\documentclass{aa}
\usepackage{amsmath}
\usepackage{txfonts}
\usepackage{epsfig,graphicx}
\usepackage{natbib}
\bibpunct{(}{)}{;}{a}{}{,}

\begin{document}
   \title{A new code for automatic determination of equivalent widths: Automatic Routine for line Equivalent widths in stellar Spectra (ARES)\thanks{ARES Webpage: http://www.astro.up.pt/$\sim$sousasag/ares/ - Table 1 is available in electronic form at the CDS via anonymous ftp to cdsarc.u-strasbg.fr (130.79.128.5) or via http://cdsweb.u-strasbg.fr/cgi-bin/qcat?J/A+A/}}


   \author{S. G. Sousa\inst{1,}\inst{2,}\inst{6}, N. C. Santos\inst{1,}\inst{3,}\inst{4}, G. Israelian\inst{5}, M. Mayor\inst{3}, M. J. P. F. G. Monteiro\inst{2,}\inst{6}         }

   \offprints{S. G. Sousa: sousasag@astro.up.pt}

   \institute{Centro de Astronomia e Astrof\'isica da Universidade de Lisboa, Observat\'orio Astron\'omico de Lisboa, Tapada da Ajuda, 1349-018 Lisboa, Portugal
\and Centro de Astrof\'isica da Universidade do Porto, Rua das Estrelas, 4150-762 Porto, Portugal
\and Observatoire de Gen\`eve, 51 Ch. des Mailletes, 1290 Sauverny, Switzerland
\and Centro de Geof\'isica de \'Evora, Col\'egio Luis Antonio Verney, \'Evora, Portugal
\and Instituto de Astrof\'isica de Canarias, 38200 La Laguna, Tenerife, Spain
\and Departamento de Matem\'atica Aplicada, Faculdade de Ci\^encias da Universidade do Porto, Portugal
     }

   \date{Received <date>; accepted <date>}


  \abstract
   {}
   {We present a new automatic code (ARES) for determining equivalent widths of the absorption lines present in stellar spectra. We also describe its use for determining fundamental spectroscopic stellar parameters.
}
   {The code is written in C++ based on the standard method of determining EWs and is available for the community. The code automates the manual procedure that the users normally carry out when using interactive routines such as the \textit{splot} routine implemented in IRAF.}
   {We test the code using both simulated and real spectra with different levels of resolution and noise and comparing its measurements to the manual ones obtained in the standard way. The results shows a small systematic difference, always below 1.5m\AA. This can be explained by errors in the manual measurements caused by subjective continuum determination. The code works better and faster than others tested before.}
   {}


\authorrunning{Sousa, S.G., Santos, N.C. et al.}

\maketitle


\section{Introduction}


When analyzing the spectrum of a star, it is possible to determine its atmospheric parameters, such as the effective temperature ($T_{\rm eff}$), surface gravity ($\log g$), and the chemical abundance for several elements \citep[][]{Gonzalez-1996, Fuhrmann-1997}. These are parameters that can be determined directly from the spectra and can be used to determine other indirect parameters.

However, this technique can be as powerful as it is time-consuming. On the one hand, measuring chemical abundances in solar type stars often implies the measurement of accurate EW's for many spectral lines. On the other hand, one of the most common, accurate methods of determing spectroscopic stellar parameters such as $T_{\rm eff}$, $\log g$, and metallicity ([Fe/H]) is based on measuring the equivalent width (EW) of weak metal lines. By using them to impose excitation and ionization equilibrium through stellar atmosphere models, we can determine the spectroscopic stellar parameters. Several authors \citep[e.g.][]{Gonzalez-1996,Gonzalez-2001,Sato-2003,Santos-2004b, daSilva-2006,Randich-2006} use similar methods to present excellent results with high precision that are very consistent.

The most common procedure for measuring EW is the use of interactive routines such as the IRAF\footnote{IRAF is distributed by National Optical Astronomy Observatories, operated by the Association of Universities for Research in Astronomy, Inc., under contract with the National Science Foundation, U.S.A.} ``splot'' routine within the {\tt echelle} package. In these cases it is usually necessary to ``manually'' find the line and to mark the position of the continuum and the position of possible additional lines in the case of blending effects. Even the inexperienced reader can imagine the time needed to analyze a spectrum in this way, taking the number of lines for each spectrum into account to determine the parameters with satisfactory accuracy. Multiplying this time by the number of stars, one can have an idea of the total time needed to study a large sample of stars for several chemical elements.

With the growing amount of incoming stellar data (e.g. planet-host star sample, comparison star samples,etc.), it becomes necessary to accelerate the analysis process. To overcome this problem, people have tried to build automated codes to measure EWs. In \citet[][]{Sousa-2006} we presented the results of one promising code for this task. From several codes found in the literature, DAOSPEC (Pancino \& Stetson in preparation) seems to be a satisfactory code for measuring EWs automatically. However, the user must pay particular attention to the choice of the input parameters for the code. Moreover, the code seems to underestimate the measurements by a significant amount.

In this paper we present a new automatic code (ARES) that can measure the EWs of absorption lines. In Sect. 2 we report the basic idea behind the procedure and the system requirements to run ARES. The ways to use the code are reported in Sect. 3, which refers to the input parameters and explains the numerical method inspired in the ``hand'' made measurements. In Sects. 4 and 5 we report the results of tests to the code both for synthetic spectra using the Sun as a reference and also for several real spectra taken with FEROS, HARPS, and UVES spectrographs. We also show a comparison between ARES and DAOSPEC EWs. In Sect. 6 we present new atomic parameters for a large sample of iron lines used for determining the stellar parameters. This sample of lines is then used to determine the stellar parameters and iron abundances for the Sun and several solar-type stars in an automatic way. The results are compared with the literature data. Section 7 summarizes the work presented in this paper.

\section{The ARES code}

\begin{figure}[b]
\begin{center}

\includegraphics[width=7cm]{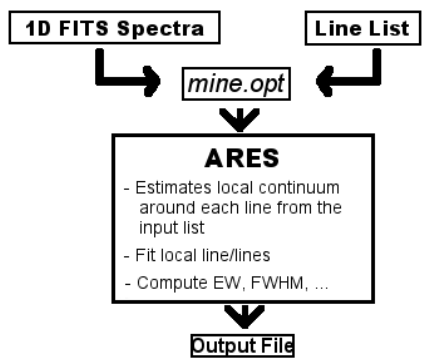}
\caption[]{Scheme for the use of ARES.}
\label{ARES_squeme}
 
\end{center}

\end{figure}


The idea behind this code is to reproduce the ``manual'' procedure of measuring the equivalent widths of absorption lines in spectra. The underlying approach is to try to ``teach'' the computer how to estimate the continuum position and to guess the number of lines (in case of blending effects) necessary for the best fit to the normalized spectrum around the line we want to measure.

The procedure is to take a one-dimensional spectrum, a list of the spectral lines to measure and a file that contains the parameters necessary for the computation. Then the code selects a region around each line, where it finds the continuum position. After this, it identifies the number and the position of the lines needed to fit the local spectrum. The fitting parameters are used to calculate the equivalent width of each line. The result is produced in a file defined by the user (see Fig. \ref{ARES_squeme}).


The ARES code was written in C++, allowing anyone to use it without having any problems with software licenses. Not only is the language free, but the libraries that this code uses are also available without any additional costs. The code can be obtained through the webpage (http://www.astro.up.pt/$\sim$sousasag/ares/), and the packages required to compile the code are:
\begin{itemize}
 \item CFITSIO (http://heasarc.nasa.gov/fitsio/fitsio.html);
 \item GCC - GNU Compiler Collection (http://gcc.gnu.org/);
 \item GSL - GNU Scientific Library (http://www.gnu.org/soft\-ware/gsl/);
 \item \textit{plotutils} (http://www.gnu.org/software/plotutils/).
\end{itemize}

All software used to run ARES, with the exception of CFITSIO, can be easily installed in any linux flavor machine, either by using \textit{yum} for Fedora Core linux or using repositories for Debian. The user can proceed by installing each software package individually by making use of the information provided on each software web page. We want to thank the Free Software Foundation (FSF) for sponsoring the GNU Project, making it possible for anyone to have access to free functional software.

\section{Numerical method and input parameters}

A series of input parameters are needed to run ARES. These parameters are given in an input file and include the 1D-fits file name (eg. IRAF format), the line-list, the output-file names, the wavelength domain to be used, the minimum interval between successive lines, the wavelength interval to be considered around each line, the minimum accepted EW of a line, calibration parameters for either the continuum determination and noise control, and the display parameters. A more thorough description of these parameters can be found in a ``README'' file that comes with the program package. Note that the input spectrum should already be reduced and corrected by the Doppler shift. We also recommend making a previously 1st-order normalization, especially for regions where the continuum has big changes.

Some other more crucial parameters, needed to fit the continuum and to smooth the spectra, are also defined here. Given their relevance,
these are described in more detail in the next sections.

\subsection{Continuum position determination}

\begin{figure}[t!]
\includegraphics[width=8cm]{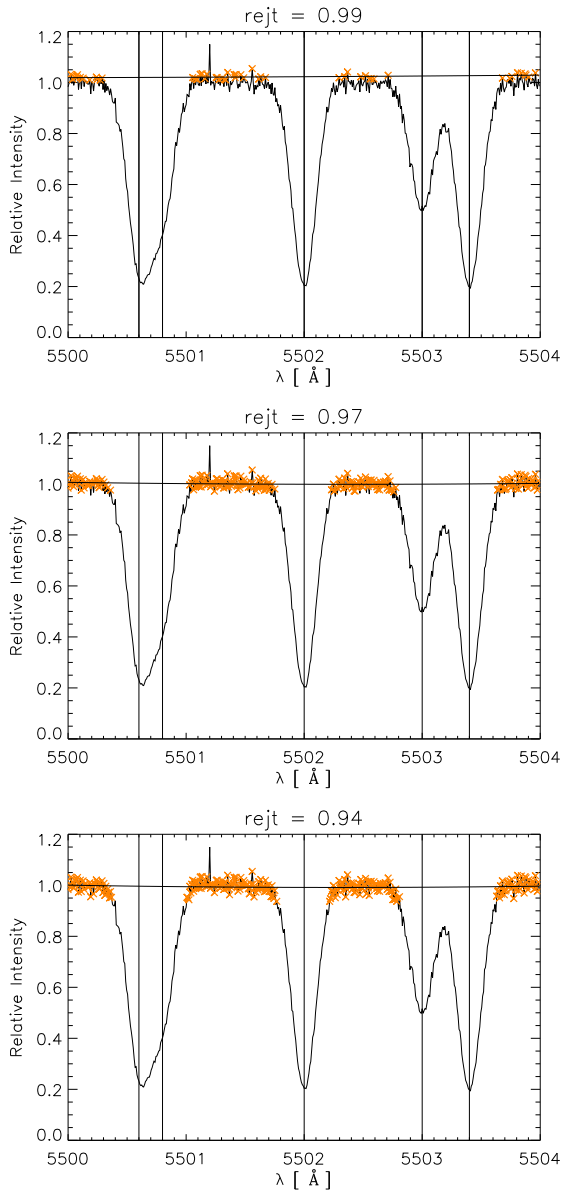}
\caption[]{A synthetic spectrum region with a signal-to-noise ratio around 50. The filled line represents the fit to the local continuum through a 2nd-order polynomial based on the points that are marked as crosses in the spectrum. We can see the variations induced by some different choises to determine the local continuum (effect of the parameter \textit{rejt}, cf. Sect. 4.1).}
\label{partree}
\end{figure}

The computation of the EW is done locally around each line that the user wants to measure. The parameter \textit{space} is defined by the user in order to choose the local interval (in Angstroms) where the calculations will be carried out. Therefore, the continuum position is also determined locally as is usually done when obtaining the EWs through the ``manual'' procedure. The continuum position is obtained iteratively by choosing points that are above the fit of a 2nd-order polynomial times a value that is defined in the parameter \textit{rejt}. This important parameter is necessary for responding to the S/N of the spectra. In this paper we recommend some values for this parameter for different kinds of noisy spectra; however, since the ``manual'' continuum placing is very subjective, we decided to leave this parameter free for the user to choose the best value.

Figure \ref{partree} illustrates the changes that the parameter \textit{rejt} introduces in the selection of the points for the continuum determination. This figure shows a region of a synthetic spectrum with an S/N around 50. Each panel is the result for a different value of the parameter \textit{rejt} where we mark the final points chosen after all the iterations have been done for the continuum determination. It is easy to see that higher values for this parameter, close to the value 1, will be advised for spectra with low noise. We present a case where the spectra has some significant noise level, therefore a value between 0.94 and 0.97 will make a more accurate continuum determination.

There is also an additional internal condition that tries to neglect possible cosmic rays/bad pixels in the spectra by taking out points that suffer a big change in the flux. This is done by checking the difference between consecutive points. If this has a high relative value (10\%), then this point is not taken into account for the continuum determination. In the synthetic spectra we introduced a bad pixel around 5501.2 \AA\ in the continuum region. It is easily seen that in all cases this pixel was ignored for the continuum determination. We stress that it is assumed that the spectrum used as input for the calculation should be reduced as much as possible and calibrated in wavelength, without cosmic rays and with some type of normalization.

Note that the position of the continuum, together with the fit of the line, can be seen if required by the user (setting \textit{plots\_flag} = 1). Then the user has the choice of continuing to run without plots if the result is satisfactory (setting \textit{plots\_flag} = 0). Alternatively the user can stop the code and reset the parameters for a better result.

\begin{figure*}[htp]
\centering
\includegraphics[width=17cm]{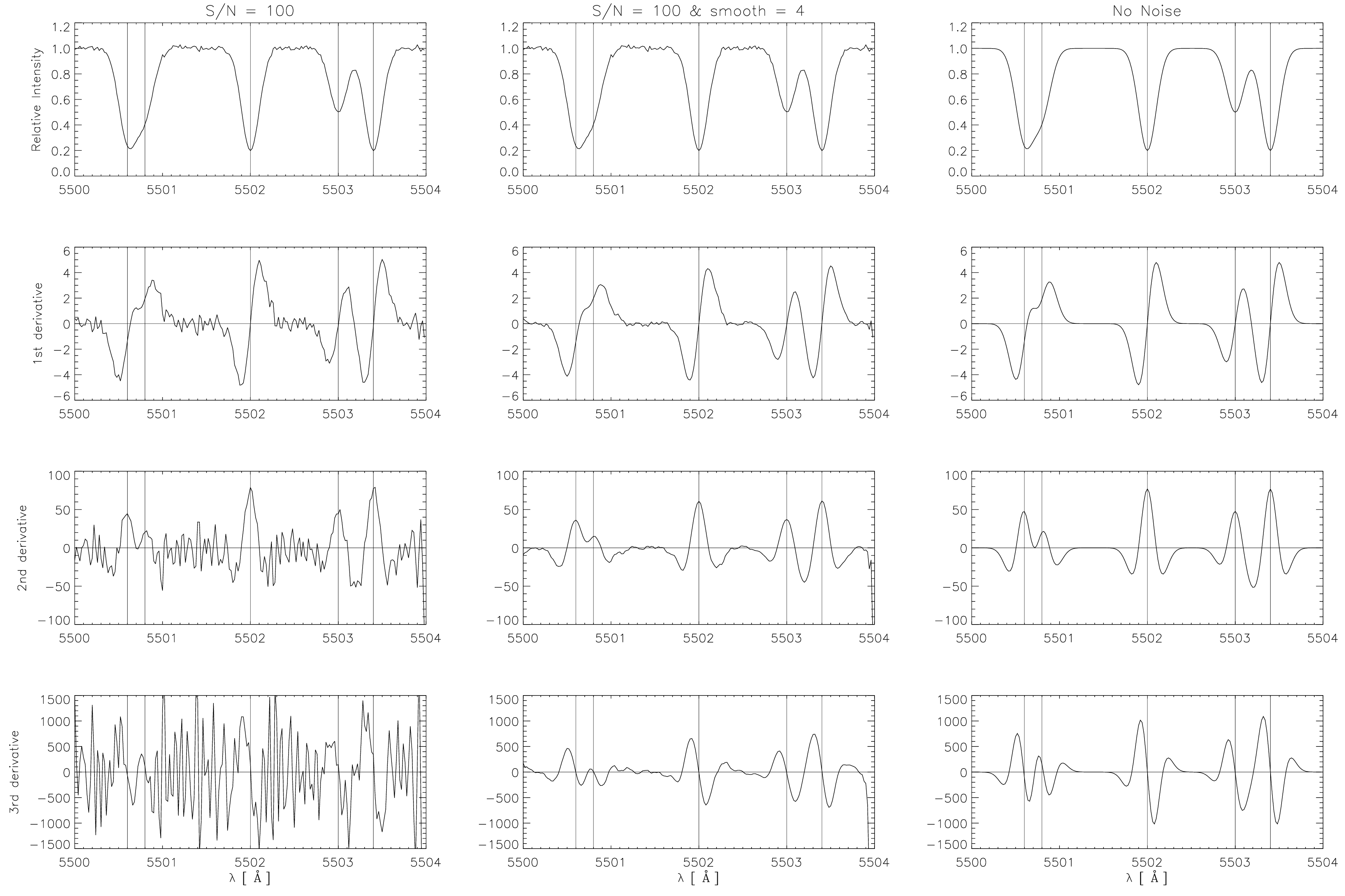}
\caption[]{A simulated normalized spectral region and the respective 1st, 2nd, and 3rd derivatives. In the right panel it is easily seen that the local maxima of the 2nd derivative, which can be found using the zeros of the 3rd derivatives, estimates the center of the lines of the simulated spectrum very well. Note that using this we can see extremely blended lines that do not present zeros on the 1st derivative. The left and middle panels show how the parameter \textit{smoothder} is useful for eliminating the noise when computating the derivatives.}
\label{dersmooth}
\end{figure*}

\begin{figure*}[htp]
\centering
\includegraphics[width=17cm]{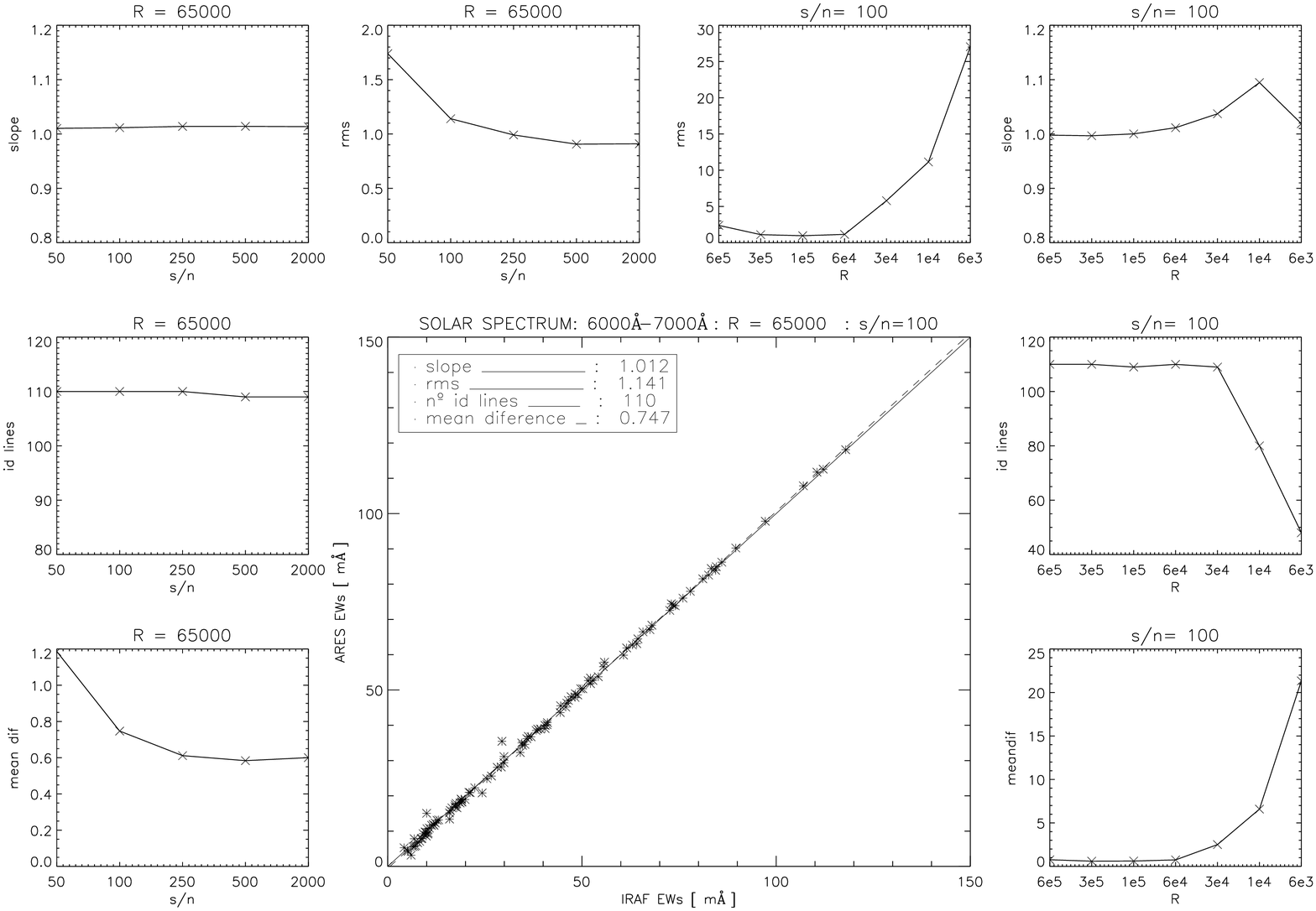}
\caption[]{ARES results for different kinds of spectra in terms of resolution and noise contamination. The larger plot shows the comparison between ARES EWs (y axes) and ``hand made'' EWs (x axes). In the box we present the slope of the linear fit to the points (dashed line), the \textit{rms}, the number of identified lines, and the mean difference in the EW of the measurements. The filled line is a 1:1 line. The small plots represent the variations in the values presented on the box for the different spectra. On the left we fixed the resolution and changed the noise levels. On the right we fixed the noise level and changed the resolution. Note that these results can be improved if a more careful choice of the input parameters is made for each spectrum type.}
\label{testes}
\end{figure*}

\subsection{Finding the lines to fit}

To calculate the EW it is necessary to fit the line with a function. Absorption lines weaker than 150 m\AA\ are described very well by a Gaussian function. In this version of the code we only fit the lines with Gaussians, but this can be easily extended to other options if necessary in the future.

When using interactive (``manual'') routines it is necessary to point the approximate position of the line peak or peaks in the case of blending lines. To enable the computer to automatically find the peaks, we use some mathematical properties of the derivatives of a function. It is well known that the zeros of the derivative of a function give us the local minima and maxima. The zeros of the 2nd derivative give us the inflection points, i.e. the points where the function changes its concavity. The local maxima of the 2nd derivative will give us the center of the absorption lines directly. The best way to find these maxima is to use the zeros of the 3rd derivative of the function. This can be seen in the right panel of Fig. \ref{dersmooth}. It is clear that the maxima of the second derivative gives us a good estimation of the position of the lines in the original spectra even for cases of strong blended lines (e.g. 5500.6 \AA\ and 5500.8 \AA\ in Fig. \ref{dersmooth}).

To determine the peak or peaks to fit the spectral line, it is necessary to obtain the first three numerical derivatives of the surroundings of the line profile.

\subsection{Dealing with noise}

Noise can be a problem, especially when determining numerical derivatives where the noise propagates very fast (left panel of Fig. \ref{dersmooth}). To overcome this problem we make use of a numerical smoothing applied to the arrays of the derivatives with a \textit{boxcar} average and a given width eliminating some of the noise. The use of such a smoothing parameter is illustrated in the middle panel of Fig. \ref{dersmooth}. The parameter \textit{smoothder} is an integer value corresponding to the width (in pixels) of the \textit{boxcar}. In the code it is only applicable to the derivatives. The smooth result of an array of values is described by the expression

\vspace{0.25cm}
$
R_i=
\begin{cases}
\frac{1}{w} \displaystyle\sum_{j=0}^{w-1} A_{i+j-w/2} & i=\frac{(w-1)}{2}\text{ ,..., }N-\frac{(w-1)}{2} \\
A_i & \text{otherwise}
\end{cases}
$
\vspace{0.25cm}
\newline where \textit{A} is the array to be smoothed, \textit{w} the width of the \textit{boxcar} to be averaged, \textit{N} the number of array elements, and \textit{R} the resulting smoothed array. We note that this parameter can be important regarding the numerical resolution of the spectra. From the tests we made using different spectra from FEROS and HARPS, which present a significant difference in spectral resolution and consequently also numerical resolution, the typical value for this parameter may change between 2 and 4.

Even using this numerical trick, noise can be a problem when identifying the lines. If the noise is high enough, the procedure can mistakenly identify lines that are very close to each other and that is only one line in reality. To overcome this possible problem the user can set the parameter \textit{lineresol} that tells the computer the minimal distance between consecutive lines. The value is introduced in Angstroms units. A typical value for this parameter is 0.1 \AA.

\begin{figure*}[htp]
\centering
\includegraphics[width=17cm]{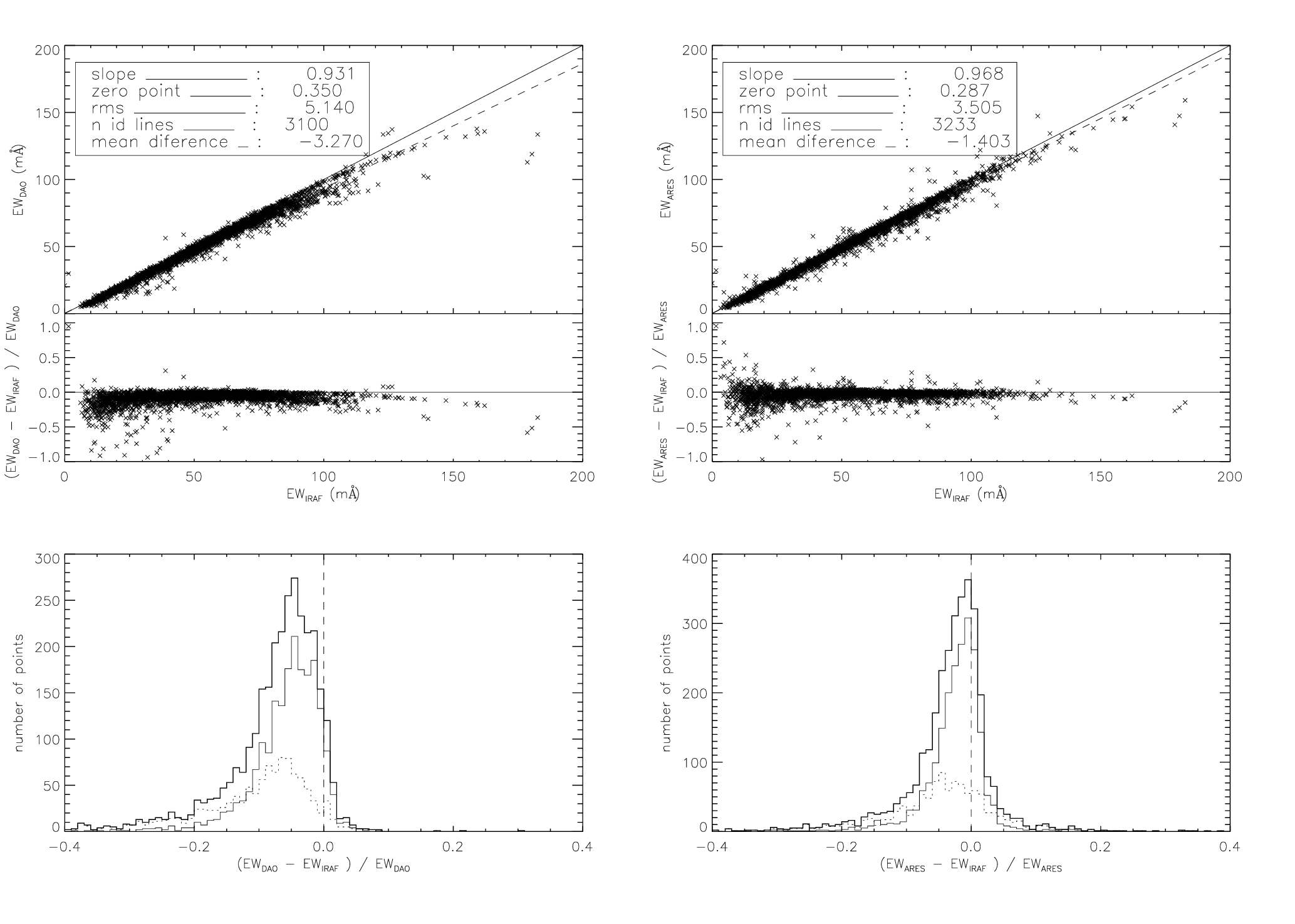}
\caption[]{Results of both ARES (right) and DAOSPEC (left) comparisons to `hand made' measurements using the same sample of FEROS spectra. The dotted histogram represents the lines that are weaker than 40 m\AA. The thin line histogram represents the lines that are stronger than 40 m\AA.}
\label{daovsares}
\end{figure*}

\section{Testing ARES: Using the solar spectrum}

To test ARES we performed the same tests that were presented in \citet[][]{Sousa-2006}. We will compare the ``hand'' made EW measurements obtained using the IRAF ``splot'' routine to the ones computed by ARES. The test will be carried out using several solar spectra with different levels of artificial noise and resolution. The original spectrum was obtained using the Kurucz Solar Atlas, and the noise was introduced using a Gaussian distribution. The artificial instrumental resolution was included using the ``rotin3'' routine in SYNSPEC\footnote{http://tlusty.gsfc.nasa.gov/index.html} \citep[][]{Hubeny-1994}.


To perform the tests we chose 114 iron lines (blended lines also included) in the region [6000\AA - 7000\AA ]. In Fig. \ref{testes} we have a set of plots that show the results of the tests. The large panel shows a comparison between the ARES EWs (\textit{y} axes) and the ``hand made measurements'' (\textit{x} axes) for a spectrum with S/N $\sim$ 100 and a resolution R $ = \lambda/\Delta\lambda \sim 65\,000$ (note that R $ \sim \lambda_c/\textit{fwhm}$, where $\lambda_c = 6500$ \AA \ and\ \textit{fwhm}=$0.1$ \AA ). In the box of this plot we indicate the slope of the linear fit to the points, the number of identified lines by ARES, the \textit{rms}, and the \textit{mean difference} of the results (in m\AA). As we can see from the plot, the points are all very close to the identity line (filled line). In this figure we also present several panels that illustrate how the values displayed in the box mentioned above change when fixing the spectral resolution (R $\sim$ 65000)-left and when fixing the noise level (S/N $\sim$ 100)-right.

In these comparisons, a few points can appear to be at a significant distance from the identity line (2-3 points in the larger plot for Fig. \ref{testes}). These offsets can appear for strongly blended lines where the fit is more difficult to compute. To check these cases, the user can see the log file that the code produces to check for bad fits in these lines.

The slopes are always very close to 1, and the \textit{rms} is always smaller than 2 m\AA\ except in the case of low resolution. The number of identified lines is very high in most of the spectra and the mean difference always show a low value in the typical spectral type. The results are generally very good, but the agreement gets worst when the spectra quality decreases, as expected. We note that the parameters used to perform this test for the different kinds of spectra were chosen in a ``fast'' way, i.e. a more careful choice of the parameters for each type of spectra is required to obtain better results. The parameters are very important, and once these are adequately chosen for a specific type of spectrum in terms of resolution and noise level, the results will be equally valid under similar conditions. We recommend the users to make a careful selection of the parameters, or to take the ones presented in the next section for reference. We also recommend the user to first use a line list composed with ``easy'' lines (isolated and well-defined) in order to see the progress of the code. The user can do this by using the plots provided by the code and choose the parameters that present the best fit to the lines and the continuum. As discussed above, the continuum position is very subjective, and each user can choose the parameters that better fit his/her demands.

\section{Testing ARES: using real stellar data}

\begin{figure}[t]
\includegraphics[width=8cm]{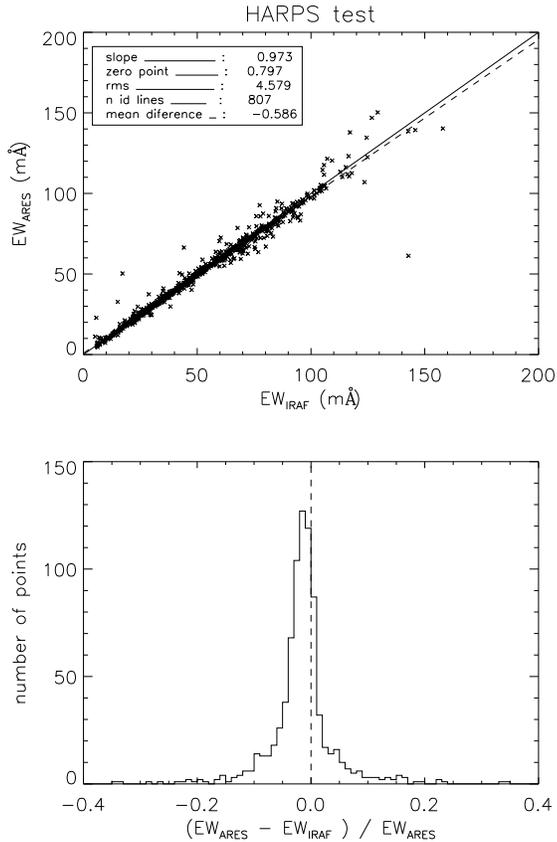}
\caption[]{Results of ARES for a sample of HARPS spectra comparing the value of the EW obtained by ARES ($EW_{ARES}$) with the ``manual'' value from IRAF ($EW_{IRAF}$).}
\label{testeharps}
\end{figure}

\begin{figure}[t]
\includegraphics[width=8cm]{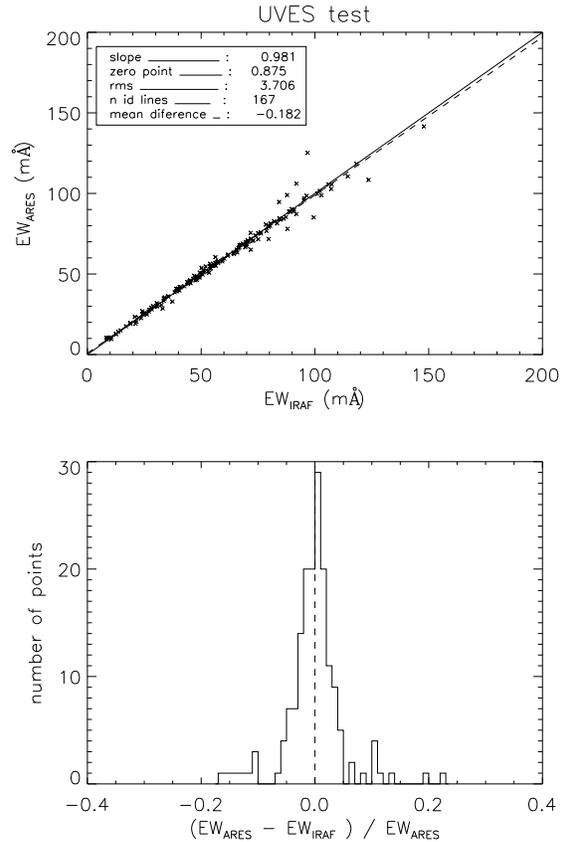}
\caption[]{Results of ARES for a sample of UVES spectra comparing the value of the EW obtained by ARES ($EW_{ARES}$) with the ``manual'' value from IRAF ($EW_{IRAF}$).}
\label{testeuves}
\end{figure}


The code has also been applied to available spectra obtained with the FEROS, HARPS, and UVES spectrographs.

\subsection{FEROS spectra; ARES vs. DAOSPEC}

A comparison to the results obtained by DAOSPEC for the same FEROS sample of spectra is also included. The spectra were obtain to characterize the planet-host stars and comparison samples of stars that was not found any planet. For details on these spectra, see the work of \citet[][]{Santos-2004b,Santos-2005} and \citet[][]{Sousa-2006}.

In Fig. \ref{daovsares} we show the results of both DAOSPEC (left) and ARES (right) versus the ``manual'' measurements. We also plot a histogram representative of the difference between the measurements to give a better idea of the dispersion and mean value of the automatic measurements. The same FEROS spectra and the same line list presented in \citet[][]{Sousa-2006} have been used. We present here the direct result from DAOSPEC without any calibration, details for the computation using DAOSPEC can be found in \citet[][]{Sousa-2006}. We also present the \textit{rms} and the \textit{mean difference} of the measurements to give a better idea of the result of the comparison.

In the case of the ARES code, since the sample is composed of high-quality spectra, with S/N $\gtrsim$ 200 and R $\sim$ 50000, the following main parameters were used: \textit{smoothder}=4, \textit{space}=2 \AA , \textit{rejt}=0.996, \textit{lineresol}=0.1 \AA, and \textit{miniline}=2m\AA. Using this parameters we derived EWs that show a small mean difference of about 1.5 m\AA\ with respect to the ``hand'' made measurements.

We suspect that part of the difference mentioned above is due to the fact that the spectra have different noise levels and that we are using the same parameters to determine EWs for all of them. This will also contribute to the dispersion presented in the plots. We cannot exclude that the subjective manual determination of the EWs may also have a significant weight in the difference presented here.

It can also be seen in Fig. \ref{daovsares}, as found before, that DAOSPEC underestimates EWs for most of the measurements. In this aspect, ARES reacts better than DAOSPEC and presents a sharper distribution, with the mean difference closer to zero. Note that the weaker lines are the ones that present higher relative error and are the ones that contribute the most to the small bump observed to the left in the histograms. This can be seen in the histogram that represents lines weaker than 40 m\AA.

Another point that favors ARES over DAOSPEC is the computation velocity. While DAOSPEC took a few hours to obtain the measurements, since we had to do it taking several small intervals for each spectra, ARES took the measurements in two minutes using the same machine. Moreover, ARES was able to identify more lines in the spectra sample than DAOSPEC.

\subsection{HARPS and UVES spectra}

In Fig. \ref{testeharps} we show the result for a sample of HARPS spectra. The main parameters used here were: \textit{smoothder}=4, \textit{space}=2 \AA , \textit{rejt}=0.993, \textit{lineresol}=0.1 \AA, and \textit{miniline}=2m\AA.
The results are very promising for all of the spectra, despite the significant variation in the noise level of the spectra in this sample (S/N $\sim$ 150-250 and R $\sim$ 115000).

In Fig. \ref{testeuves} we show the result for a small sample of four UVES spectra. The main parameters used here were: \textit{smoothder}=4, \textit{space}=2 \AA, \textit{rejt}=0.999, \textit{lineresol}=0.1 \AA\ and \textit{miniline}=2m\AA.
The four spectra, used here to test the code on UVES spectra, were of excellent quality with a very high S/N $\gtrsim$ 200 and R $\sim$ 100000. Therefore the value of the \textit{rejt} parameter is also high.

The code seems to work very nicely with these spectra, using the parameters presented here.


\section{Stellar parameters}
\subsection{Using ARES EWs}

\begin{figure*}[htp]
\centering
\includegraphics[width=17cm]{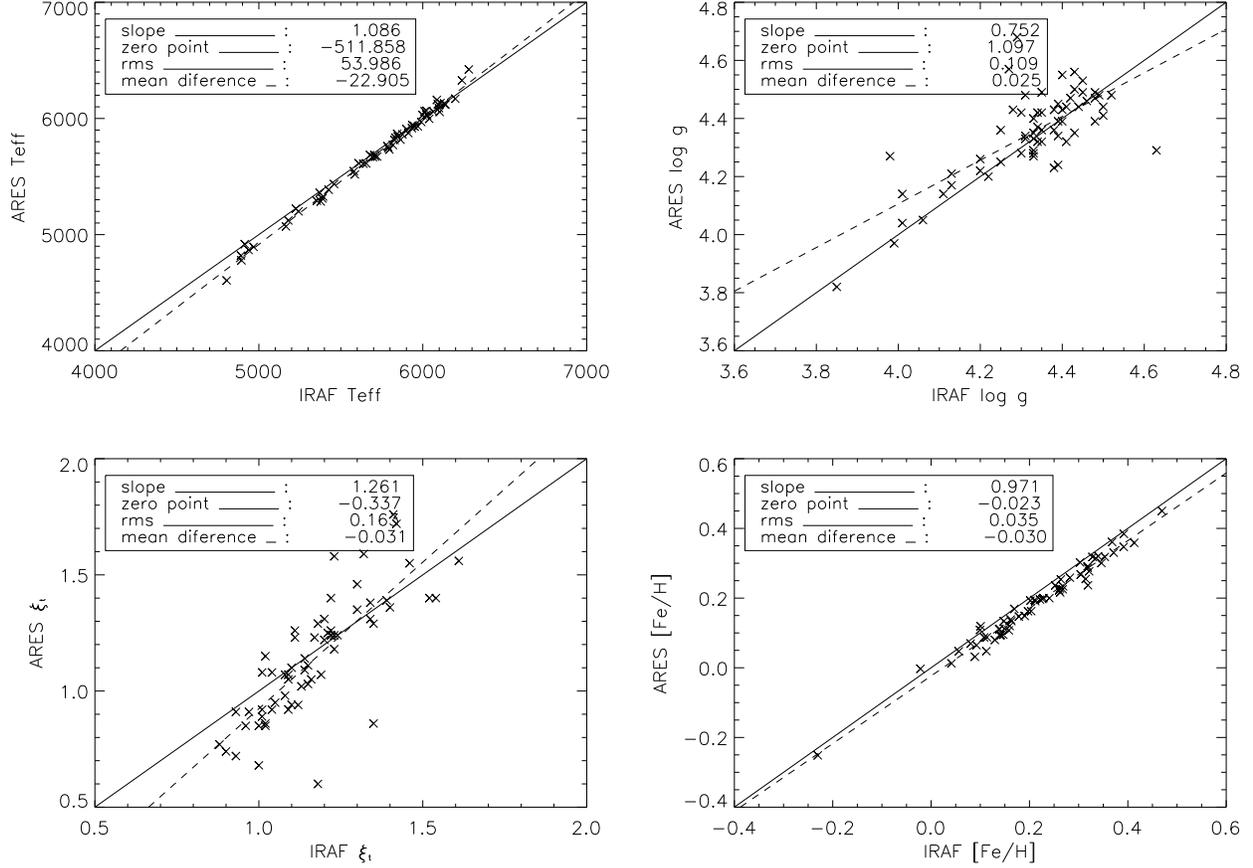}
\caption[]{Comparison of the manual obtained spectroscopic parameters to the obtained automatically through ARES code}
\label{parameters}
\end{figure*}

We re-determined the stellar parameters for the FEROS sample of stars presented in \citet[][]{Sousa-2006} to compare the results. The EWs were measured with ARES, with the same parameters as described before for this FEROS sample. After this, the same procedure was used, using the same iron line list for a spectroscopic analysis done in LTE using the 2002 version of the code MOOG\footnote{The source code of MOOG2002 can be downloaded at http://verdi.as.utexas.edu/moog.html} \citep[][]{Sneden-1973} and a grid of Kurucz Atlas plane-parallel model atmospheres \citep[][]{Kurucz-1993}.

Figure \ref{parameters} shows the comparison of the parameters obtained using ARES versus the parameters obtained using the manual EW measurements. We can see from the figure that the parameters are all very close to the identity line. The effective temperature has a mean difference of 22 K and the metallicity has a mean difference of 0.03 dex. These differences can be explained by the small difference shown before of 1.5 m\AA\, presented in the EWs measurements (see Sect. 6.1).

\subsection{Extended list of iron lines}

We built a code that presents promising results in the automatic EW determination. This was the first step in the quest for a total automatic procedure to determine spectroscopic stellar parameters. The second step is to determine a way to obtain the stellar parameters with high accuracy. The idea is to use a significant number of iron lines to increase the statistical strength of the determined stellar parameters. 

\begin{table}[t]
\caption[]{Sample Table: Atomic parameters and measured solar equivalent widths for
\ion{Fe}{ii} and \ion{Fe}{i} lines.}
\begin{tabular}{lccc}
\hline
\noalign{\smallskip}
$\lambda$ (\AA) & $\chi_{l}$ & $\log{gf}$ & EW$_{\sun}$ (m\AA)\\
\hline
  4000.01 & 2.83 & $-$3.687 & 7.3 \\
  4007.27 & 2.76 & $-$1.666 & 87.7 \\
  4010.18 & 3.64 & $-$2.031 & 35.0 \\
  4014.27 & 3.02 & $-$2.330 & 47.3 \\
  4080.88 & 3.65 & $-$1.543 & 58.2 \\
  4114.94 & 3.37 & $-$1.720 & 61.2 \\
  4124.49 & 3.64 & $-$2.071 & 33.7 \\
  4126.86 & 2.85 & $-$2.638 & 41.3 \\
  4139.93 & 0.99 & $-$3.423 & 80.4 \\
  ...     & ...  &  ...     & ...  \\
\hline
\end{tabular}
\label{tab1}
\end{table}

We follow the procedure presented in \citet[][]{Santos-2004b} for derivating stellar parameters and metallicities. It is based on the EW of 39 \ion{Fe}{i} and 16 \ion{Fe}{ii} weak lines, by imposing excitation and ionization equilibrium. Presently the number of lines are relatively small since users are restricted to a few well-defined iron lines in the stellar spectrum. The idea is to increase this number of lines, giving the possibility of decreasing the errors involved in the determination of the stellar parameters.

Using the Vienna Atomic Line Data-base (VALD: \citet[][]{Piskunov-1995, Kupka-1999,Ryabchikova-1999} - http://www.astro.uu.se/htbin/vald), we requested all the \ion{Fe}{i} and \ion{Fe}{ii} from 4000\AA\ to 9000\AA. Then using the Kurucz solar atlas, we searched all the iron lines allowing a good measurement. Then the $\log gf$ values were obtained. For this a differential analysis to the Sun was used; i.e. the $\log gf$ values were determined using the lines present in the Sun. We computed from an inverted solar analysis using EW measured from the Kurucz Solar Atlas and a Kurucz grid model for the Sun \citep[][]{Kurucz-1993} having T$_{\mathrm{eff}} = 5777 K$, $\log{g} = 4.44 $ dex, $\xi_t = 1.00$ $km s^{-1}$, and $log (Fe)$ = 7.47. The new values of $\log gf$ are presented in Table \ref{tab1}.

As a test, we computed the solar parameters and iron abundances based on iron EW measured using a solar spectrum taken with the HARPS spectrograph using the Ceres asteroid (Collection of HARPS solar spectra - http://www.ls.eso.org/lasilla/sciops/3p6/harps/monitoring/ sun.html). The resulting parameters were $T_{\rm eff}=5762 \pm 12$ , $\log g = 4.44 \pm 0.08$, $\xi_t = 0.94 \pm 0.01$, and [Fe/H] $= 0.0 \pm 0.02$, close to the ``expected'' solution.

\begin{figure}[htp]
\includegraphics[width=8cm]{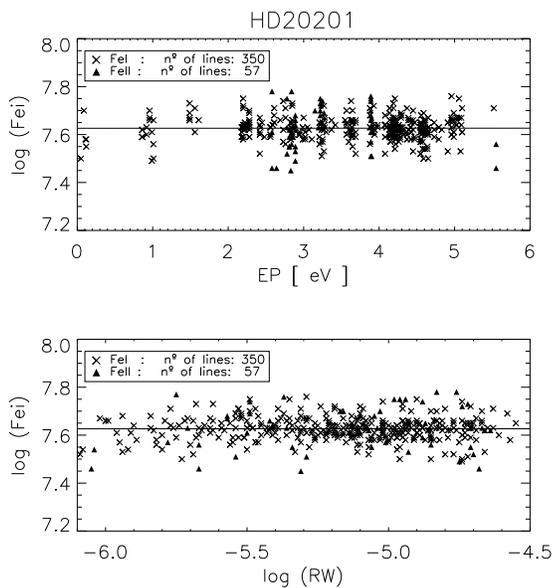}
\caption[]{Results for the fit of the abundances for determining the stellar parameters for the FEROS spectra of HD20201. The crosses represent the \ion{Fe}{i}, and the filled triangles the \ion{Fe}{ii}.}
\label{HD20201}
\end{figure}

We also show results for a few stars in the FEROS and HARPS sample. The stellar parameters obtained for these stars were obtained using ARES to measure EW with this new line list. The procedure is the same as for the short line list and the stellar parameters are presented in Table \ref{tab:parameters} where we can compare them to previous results added from the literature. The result of imposing the excitation and ionization equilibrium can be seen in Fig. \ref{HD20201} for the HD20201 star. The results are compatible with the previous ones obtained using the short line list, but since we have more lines, this is statistically more significant.

\begin{table*}
\caption[]{Stellar parameters for some test stars}

\begin{tabular}{lcccrccc}
\hline
\hline
HD number    & T$_{\mathrm{eff}}$ [K] & $\log{g}_{spec}$ [cm\,s$^{-2}$] & $\xi_{\mathrm{t}}$ [km\,s$^{-1}$]& \multicolumn{1}{c}{[Fe/H]} & N(\ion{Fe}{i},\ion{Fe}{ii}) & $\sigma$(\ion{Fe}{i},\ion{Fe}{ii}) & Ref.$\dagger$  \\
\hline

FEROS spectra:& & & & & & \\
\hline
\object{HD\,20201}   &6064\ $\pm$\ 34 &  4.43\ $\pm$\  0.02& 1.17\ $\pm$\  0.05 &0.19\ $\pm$\  0.05&38,15& 0.04, 0.06 & [1]\\
		 &6084\ $\pm$\ 18 &  4.57\ $\pm$\  0.03& 1.16\ $\pm$\  0.02 &0.16\ $\pm$\  0.01&350,57& 0.05, 0.09 & [2]\\	

\object{HD\,33214}   &5180\ $\pm$\ 74 &  4.40\ $\pm$\  0.11& 1.10\ $\pm$\  0.10 &0.17\ $\pm$\  0.08&39,14& 0.08, 0.13 & [1]\\
		 &5265\ $\pm$\ 31 &  4.48\ $\pm$\  0.08& 1.10\ $\pm$\  0.06 &0.19\ $\pm$\  0.02&348,51& 0.09, 0.24 & [2]\\	

\object{HD\,36553}   &6103\ $\pm$\ 46 &  3.85\ $\pm$\  0.03& 1.61\ $\pm$\  0.07 &0.41\ $\pm$\  0.06&37,15& 0.05, 0.09 & [1]\\
		 &6153\ $\pm$\ 23 &  4.06\ $\pm$\  0.04& 1.67\ $\pm$\  0.02 &0.39\ $\pm$\  0.02&349,55& 0.07, 0.12 & [2]\\	

\object{HD\,7570}    &6198\ $\pm$\ 39 &  3.49\ $\pm$\  0.02& 1.40\ $\pm$\  0.07 &0.24\ $\pm$\  0.05&38,15& 0.04, 0.06 & [1]\\
		 &6204\ $\pm$\ 16 &  4.55\ $\pm$\  0.03& 1.31\ $\pm$\  0.02 &0.21\ $\pm$\  0.01&350,59& 0.05, 0.10 & [2]\\	

\object{HD\,7727}    &6131\ $\pm$\ 41 &  3.34\ $\pm$\  0.02& 1.18\ $\pm$\  0.07 &0.16\ $\pm$\  0.05&38,15& 0.05, 0.07 & [1]\\
		 &6169\ $\pm$\ 18 &  4.55\ $\pm$\  0.03& 1.26\ $\pm$\  0.02 &0.12\ $\pm$\  0.01&327,57& 0.05, 0.10 & [2]\\	

\hline
HARPS spectra:& & & & & & \\
\hline

\object{HD\,102117}  &5672\ $\pm$\ 22 &  4.27\ $\pm$\  0.03& 1.05\ $\pm$\  0.02 &0.30\ $\pm$\  0.02&37, 8& 0.03, 0.03 & [3]\\
		 &5671\ $\pm$\ 18 &  4.24\ $\pm$\  0.03& 1.06\ $\pm$\  0.02 &0.28\ $\pm$\  0.01&286,56& 0.05, 0.10 & [2]\\	

\object{HD\,212301}  &6256\ $\pm$\ 28 &  4.52\ $\pm$\  0.02& 1.43\ $\pm$\  0.06 &0.18\ $\pm$\  0.02&36, 9& 0.03, 0.01 & [1]\\
		 &6273\ $\pm$\ 21 &  4.64\ $\pm$\  0.04& 1.32\ $\pm$\  0.02 &0.16\ $\pm$\  0.02&278,53& 0.05, 0.13 & [2]\\	

\object{HD\,4308}    &5685\ $\pm$\ 12 &  4.49\ $\pm$\  0.02& 1.08\ $\pm$\  0.03 &-0.31\ $\pm$\  0.01&35, 8& 0.01, 0.02 & [1]\\
		 &5675\ $\pm$\ 13 &  4.43\ $\pm$\  0.02& 0.91\ $\pm$\  0.02 &-0.33\ $\pm$\  0.02&293,56& 0.04, 0.08 & [2]\\	

\object{HD\,69830}   &5385\ $\pm$\ 19 &  4.37\ $\pm$\  0.03& 0.80\ $\pm$\  0.03 &-0.05\ $\pm$\  0.01&35, 8& 0.03, 0.02 & [1]\\
		 &5410\ $\pm$\ 17 &  4.35\ $\pm$\  0.04& 0.83\ $\pm$\  0.02 &-0.07\ $\pm$\  0.01&284,55& 0.05, 0.15 & [2]\\

\hline
\end{tabular}
\\ $\dagger$ The instruments used to obtain the spectra were: [1] \citet[][]{Sousa-2006}, [2] this paper, [3] \citet[][]{Santos-2005};
\label{tab:parameters}
\end{table*}

\subsection{A more accurate analysis}

An alternative procedure that can improve the accuracy of the stellar parameters is being considered. A differential spectroscopic analysis \citep[][]{Laws-2001} between the Sun and a solar-type star can eliminate intrinsic errors in the determination of the stellar parameters. If we use the same spectrograph to obtain both the Sun and the target star spectra, and also use the same method to determine the EWs, it is possible that, when comparing the individual line abundances for the both star, intrinsic errors either on the EW measurements or in the $\log gf$ values can be eliminated. Using ARES for the EW measurements, we are now able to apply this procedure to a large sample of lines.

Another alternative option that can also improve the analysis is to compute new $\log gf$ values based on ARES measurements for the solar spectrum obtained with the same spectrograph (same instrumental and numerical resolution) and noise levels. In this way, a systematic error in the EW measurement made by ARES will be taken into acount in the respective new solar $\log gf$ value.

\section{Summary}

We present ARES, a new automated code for measuring EWs. We describe the method used by this new code and also the requirements for its use.

The results of the test performed with ARES concludes that this code seems to perform very well for several types of spectra. There is a small difference between ARES and manual measurements that can in part be explained by the intrisic errors on the subjective determination of the continuum level. The general results of the EW measurements using real data show excellent results, especially in high-resolution spectra.

We also present a new large line list of iron lines for determining the stellar parameters to be used in an automatic procedure. The result for the Sun and for the stars of the FEROS and HARPS samples shows that the combination of ARES with this large line list can be a powerful tool for the statistical precision of determining automated stellar parameter.

\begin{acknowledgements}
S.G.S and N.C.S. would like to acknowledge the support from the Funda\c{c}\~ao para a Ci\^encia e Tecnologia (Portugal) in the form of two fellowships and grants POCI/CTE-AST/56453/2004 and POCTI/CTE-AST/57610/2004 from POCTI, with funds from the European program FEDER.
\end{acknowledgements}

\bibliographystyle{aa}
\bibliography{sousa_bibliography.bib}

\end{document}